\RequirePackage{lineno}
\documentclass[preprint]{aastex} 

\slugcomment{Accepted for publication in ApJ, Jan 3, 2014}

\shorttitle{Reconnection in plasma turbulence}
\shortauthors{Haynes et al.}

\usepackage{comment} 

\begin{document}


\title{Reconnection and electron temperature anisotropy in sub-proton scale plasma turbulence}


\author{C. T. Haynes\altaffilmark{1}\altaffilmark{2}, D. Burgess\altaffilmark{2}, and E. Camporeale\altaffilmark{3}}
\altaffiltext{1}{c.t.haynes@qmul.ac.uk}
\altaffiltext{2}{School of Physics and Astronomy, Queen Mary University of London, Mile End Road, London E1 4NS, UK}
\altaffiltext{3}{T-5 Applied Mathematics and Plasma Physics, Los Alamos National Laboratory, Los Alamos, NM 87545, USA}


\begin{abstract}
Turbulent behavior at sub-proton scales in magnetized plasmas is important for a full understanding of the energetics of astrophysical flows such as the solar wind. We study the formation of electron temperature anisotropy due to reconnection in the turbulent decay of sub-proton scale fluctuations using two dimensional, particle-in-cell (PIC) plasma simulations with realistic electron-proton mass ratio and a guide field out of the simulation plane. A fluctuation power spectrum with approximately power law form is created down to scales of order the electron gyroradius. In the dynamic magnetic field topology, which gradually relaxes in complexity, we identify the signatures of collisionless reconnection at sites of X-point field geometry. 
The reconnection sites are generally associated with regions of strong parallel electron temperature anisotropy. The evolving topology of magnetic field lines connected to a reconnection site allows spatial mixing of electrons accelerated at multiple, spatially separated reconnection regions. This leads to the formation of multi-peaked velocity distribution functions with a strong parallel temperature anisotropy. In a three-dimensional system, supporting the appropriate wave vectors, the multi-peaked distribution functions would be expected to be unstable to kinetic instabilities, contributing to dissipation. The proposed mechanism of anisotropy formation is also relevant to space and astrophysical systems where the evolution of the plasma is constrained by linear temperature anisotropy instability thresholds. The presence of reconnection sites leads to electron energy gain, nonlocal velocity space mixing and the formation of strong temperature anisotropy; this is evidence of an important role for reconnection in the dissipation of turbulent fluctuations.
\end{abstract}






\section{Introduction}

Turbulence is almost certainly ubiquitous in astrophysical plasma flows, and it is crucial for a full understanding of energetic particle propagation and the transport and dissipation of energy. Recent advances in the understanding of astrophysical plasma turbulence have followed from the interpretation of in situ measurements of space plasmas such as the solar wind and magnetosheath. For the solar wind at MHD scales, a nonlinear active turbulent cascade operates which is dominated by Alfv\'enic fluctuations \citep[e.g., review by][]{Horbury:2005}. The power law scaling of fluctuations in the inertial range (with periods of hours to tens of seconds) is approximately Kolmogorov with a power law $\sim f^{-5/3}$, although the turbulence is also characterized by a number of other properties such as anisotropy and intermittency. For a recent review of solar wind turbulence properties in the inertial range see \citet{bruno:2013}.

At higher frequencies, in a collisionless plasma such as the solar wind, one might expect that the viscous dissipation of hydrodynamic turbulence is replaced by processes operating at particle kinetic scales, such as cyclotron or Landau damping. Indeed, in the solar wind, at frequencies corresponding to the characteristic proton scales, the magnetic fluctuation spectrum shows a break, above which it steepens with a spectral index varying between -2 and -4 \citep{leamon:1998,Bale:2005,Smith:2006}, indicating that ion kinetic processes are at work. At 1 au the frequencies corresponding to the Doppler shifted proton gyroradius and proton inertial length are usually close to each other, making it difficult to infer the appropriate scaling. However, the variation of the ion-scale break frequency with radial distance from the Sun indicates that it is related to the proton inertial length, rather than the proton gyroradius \citep{Bourouaine:2012}. A similar break of the spectrum at ion scales is also characteristic for ionospheric conditions. \citep{Kelley:1982, Hysell:1994}

Dissipation processes at the proton kinetic scale are clearly important, but recent observational work has addressed the question whether solar wind turbulent fluctuations extend down to electron scales (the electron thermal gyroradius $\rho_e \sim 1$ km at 1 au) \citep{Sahraoui:2009,Kiyani:2009, Alexandrova:2009}. Several simulation and theoretical studies have investigated the nature of collisionless plasma turbulence down to electron scales \citep{camporeale:2011,chang:2011,howes:2011,gary:2012}. These studies are concerned with properties such as the scaling of the turbulent spectrum and wave-vector anisotropy, which are vital to the problems of energy dissipation and particle acceleration in astrophysical plasmas. There is some controversy over the scaling of the turbulence at sub-proton scales, due in part to the difficulty of making observations with the required resolution and sensitivity. \citet{Alexandrova:2009} found evidence that the solar wind fluctuation spectrum ends with an exponential cut-off at electron kinetic scales. \citet{sahraoui:2013} (ApJ in press), on the other hand, showed that a spectral power break is frequently seen for events when the signal to noise ratio is sufficiently large. In these cases the spectra above the frequency corresponding to $\rho_e$ steepen with a slope of $\sim -4$ on average. This is roughly in agreement with the results of 2-D and 3-D PIC simulations \citep{camporeale:2011,chang:2011,gary:2012}.

In terms of the nature of the turbulent fluctuations at sub-proton scales two linear wave modes have been suggested as relevant for the solar wind where the amplitudes  are (on average) small at small scales: namely, kinetic Alfv\'en waves (KAW) and whistler waves. \citet{salem:2012} and \citet{chen:2013} discuss the expected observational signatures in these two cases, and present evidence suggesting that KAW dominate the turbulent fluctuations to well below the proton scale. In the weak turbulence scenario damping of fluctuations in these short scale wave modes, via particle heating, provides the dissipation required to terminate the turbulent cascade. This approach downplays the role of coherent structures (e.g., discontinuities) and the magnetic field topology within the plasma. However, the role of magnetic reconnection within turbulence, and vice versa, has been studied for many years, both by means of MHD simulations \citep[e.g.,][]{Matthaeus:1986} and observationally \citep[e.g.,][]{Eastwood:2009}. Magnetic reconnection converts magnetic energy to thermal and kinetic energy, and relaxes the complexity of the magnetic field topology, thus it potentially has an important role in the evolution and dissipation of turbulence. For example, \citet{Servidio:2010} has studied, using 2-D MHD simulations, the statistics of reconnection sites that evolve self-consistently within fully developed MHD turbulence. \citet{servidio:2011} have made the case that reconnection should be viewed as an intrinsic element of plasma turbulence: ``it would be difficult to envision a turbulent cascade that proceeds without change of magnetic topology.'' In the usual Kolmogorov-like turbulent cascade energy is considered (both magnetic and kinetic) as being transported from large to small scales, although there can be an inverse cascade in the opposite direction. The direction of energy transfer depends on nonlinear wave-wave interactions and the length scale at which energy is injected. In the heliosphere the energy injection scale is usually considered to be large, such as in the motion of magnetic field lines anchored to the turbulent solar surface, creating large scale waves, such as Alfv\'en waves \citep[e.g.,][]{Perez:2013}. If the turbulent cascade also transports, or creates, magnetic field topological complexity at small scales, the question arises as to what extent reconnection at kinetic scales affects or contributes to turbulent dissipation. 

Observations of reconnection within turbulence, away from large-scale boundaries such as the magnetopause, are rare. Some evidence has been presented of reconnection events associated with magnetic nulls in the large amplitude turbulence in the magnetosheath, downstream of the quasi-parallel part of the Earth's bow shock \citep{retino:2007}. The observations were interpreted in the context of thin current sheets between magnetic islands, and a reconnection geometry with Hall currents \citep[Fig.~1,][]{retino:2007}. Other observations have offered indirect evidence of reconnection contributing to solar wind dissipation \citep{osman:2011,Bourouaine:2012}. There is also strong evidence for quasi-steady reconnection associated with solar wind discontinuities \citep{gosling:2008}.

In this paper, we present results of fully kinetic plasma simulations with realistic mass ratio and plasma parameters (such as the ratio of plasma frequency to gyrofrequency) which are relevant to the solar wind and magnetosheath. We focus on the turbulent relaxation at sub-proton scales, and the resulting electron flows and velocity distributions. We show that reconnection sites within turbulence can be responsible for strong electron temperature anisotropy via a velocity space mixing mechanism. The development of electron temperature anisotropy is well documented for individual reconnection sites in isolated current sheets, such as those observed in the magnetotail, and is explained by a model of passing and trapped electrons \citep[e.g.,][]{Egedal:2012}. Electron temperature anisotropy has also previously been reported in PIC simulations of turbulence \citep{camporeale:2011,Karimabadi:2013}. Based on analysis of the electron dynamics, we describe a mechanism for creating electron temperature anisotropy that requires multiple magnetic reconnection sites within the turbulent field.  Electrons are accelerated in the reconnection electric field, but in magnetic turbulence the sense of reconnection (and direction of the reconnection electric field) varies between the reconnection sites.  The topology of the magnetic field linking the different reconnection sites allows them to also act as mixing zones for the accelerated particles. This leads to the formation of multi-peaked distributions in electron velocity space, which may be a source of further waves and particle coupling via instabilities. The reconnection sites within the turbulence lead to electron energy gain, nonlocal velocity space mixing and the formation of strong temperature anisotropy, all of which may contribute to the dissipation of turbulent fluctuations at sub-proton scales.

In this scenario it is also important to remember that the expanding solar wind develops non-Maxwellian velocity distribution functions (VDF), but on a statistical basis the frequency distribution of observations in a parameter space such as $T_\|/T_\perp$ against $\beta_\|$ is constrained within boundaries related to the marginal growth of linear instabilities. Typically the relevant constraints derive from temperature anisotropy instabilities such as the cyclotron, fire hose and mirror instabilities \citep{Hellinger:2006}, and the VDF evolves in a competitive balance between the effects of solar wind expansion, growth of linear instabilities and Coulomb collisions \citep{Hellinger:2008,Matteini:2012}. Although most work has concentrated on proton parameter constraints, similar effects are also seen for electrons \citep{stverak:2008}. The role of such linear instability parameter constraints in solar wind turbulence is still unclear. In this paper we argue that an additional driver for the electron temperature anisotropy in an expanding plasma flow might be magnetic reconnection occurring as an element within turbulence.

\section{Methodology}
We use the particle-in-cell (PIC) code Parsek2D \citep{markidis:2009} based on the implicit moment method for time advance of the electromagnetic fields, and a predictor-corrector method for the particle mover. The implicit method allows larger time steps and cell sizes compared with explicit PIC methods, which are usually constrained (for numerical stability) by the condition $ \omega_{pe} \Delta t < 2$, where $\Delta t$ is the time step, and $\omega_{pe}$ is the electron plasma frequency. Also Parsek2D allows a relaxation of the Courant-Friedrichs-Lewy (CFL) condition $c\Delta t/\Delta x < 1 $, where $c$ is the speed of light and $\Delta x$ is the cell size. The time step $\Delta t=0.05\Omega_e^{-1}$, where $\Omega_e$ is the electron gyrofrequency (so that the electron cyclotron motion is fully resolved), and the cell size $\Delta x = \Delta y \sim 17 \lambda_D$, where $\lambda_D$ is the Debye length. The code is two dimensional in the $x$-$y$ plane but retains all three vector components for velocities and fields. The electron-proton plasma is initially loaded with a uniform, isotropic Maxwellian distribution. The simulation box is $200 \times 200$ cells, with periodic boundary conditions and 6400 simulation particles per cell for each species. This large number of particles reduces the statistical particle noise so that the dynamic range in Fourier space is large enough to resolve the formation of a turbulent cascade. The simulation box is sized to resolve wave vectors ranging from $k\rho_e = 0.1$ to $k\rho_e = 10$, where $k$ is wave vector and $\rho_e$  the thermal electron gyroradius. The box length is about 1 ion inertial length, and the electron  gyro-motion is resolved with $\sim 3$ cells per electron gyroradius (based on the initial guide field strength). We use plasma parameters which are appropriate to the solar wind and magnetosheath: the ion plasma frequency to ion cyclotron frequency ratio $\omega_{pi}/\Omega_{i} \sim $ 1650, and the ion to electron mass ratio is physical with $m_i/m_e = 1836$. The ions and electrons are initialized to the same temperature $\beta_e = \beta_i = 0.5$. The simulation was run until $t = 200\Omega_e^{-1}$. Unless quoted otherwise, simulation results are shown in Gaussian CGS units with the following normalizations: Velocities are normalized to the speed of light, time is normalized to $10\omega_{pe}^{-1}$, and charge per unit mass is normalized to the proton charge per unit mass. Temperatures are the variance of velocities in each simulation cell. Initial particle density is assumed to be equivalent to 10 particles per $cm^3$.

Similar to the method of \citet{camporeale:2011}, we initialize the simulation with a background magnetic field $\mathbf{B_0}$ and add random long wavelength magnetic field fluctuations, but here the background field is in the out-of-plane $z$ direction. The intention is to provide an initial input of energy at low values of $k$ with properties mimicking a turbulent field, and then follow the decay of this initial perturbation and the development of power at larger wave numbers. The magnetic field is initialized with random fluctuations in all three components for wave vectors $k_x=2\pi m/L_x$ and $k_y=2\pi n/L_y$ for $m=-3,\ldots,3$ and $n=-3,\ldots,3$. We do not impose any spectral slope on the initial fluctuations. The initial electric field is zero, but the abrupt perturbation of the magnetic field acts to initialize the self-consistent evolution of the turbulent decay after a short period at the start of the simulation. This method emphasizes the random nature of turbulence, and in particular has the advantage that no particular linear modes are assumed dominant. Other methods of initializing the decay of turbulent fluctuations are possible, such as initial equilibria \citep{Karimabadi:2013}, Alfv\'enic-like fluctuations \citep{camporeale:2011}, or superposition of linear modes \citep{chang:2011}. The requirement to resolve the development of a turbulent cascade means that the initial perturbation has to be relatively large, and a value $\delta B/B_0 = 1$ is used here. The choice of a configuration with the background field perpendicular to the simulation plane does not support $k_\|$ wave vectors (at least on average), but does favor magnetic field line topologies with islands and X-points with a guide field. 

We identify magnetic field X-points as potential reconnection sites using the technique described by \citet{servidio:2009}. The vector potential $A_z$ is computed from the magnetic field; contours of constant $A_z$ represent magnetic field lines in the $x$-$y$ simulation plane. For each cell we calculate the Hessian matrix for $A_z$ and its eigenvalues. If the eigenvalues are of opposite sign, then the location is a saddle point, and if the gradient of $A_z$ is also zero then this is a potential location for reconnection. Since the simulation is discrete in space a threshold is used to indicate a possible zero gradient. This method may find multiple locations around a single reconnection point, and in this case the cell with the lowest gradient of $A_z$ is taken to be the actual center. Additionally we use in-plane field lines to confirm the calculated positions of reconnection sites, and in order to compare the field line geometry and motion with the electron bulk flow velocities.

\section{Results}

\begin{figure} 
\includegraphics[scale=0.5,angle=0]{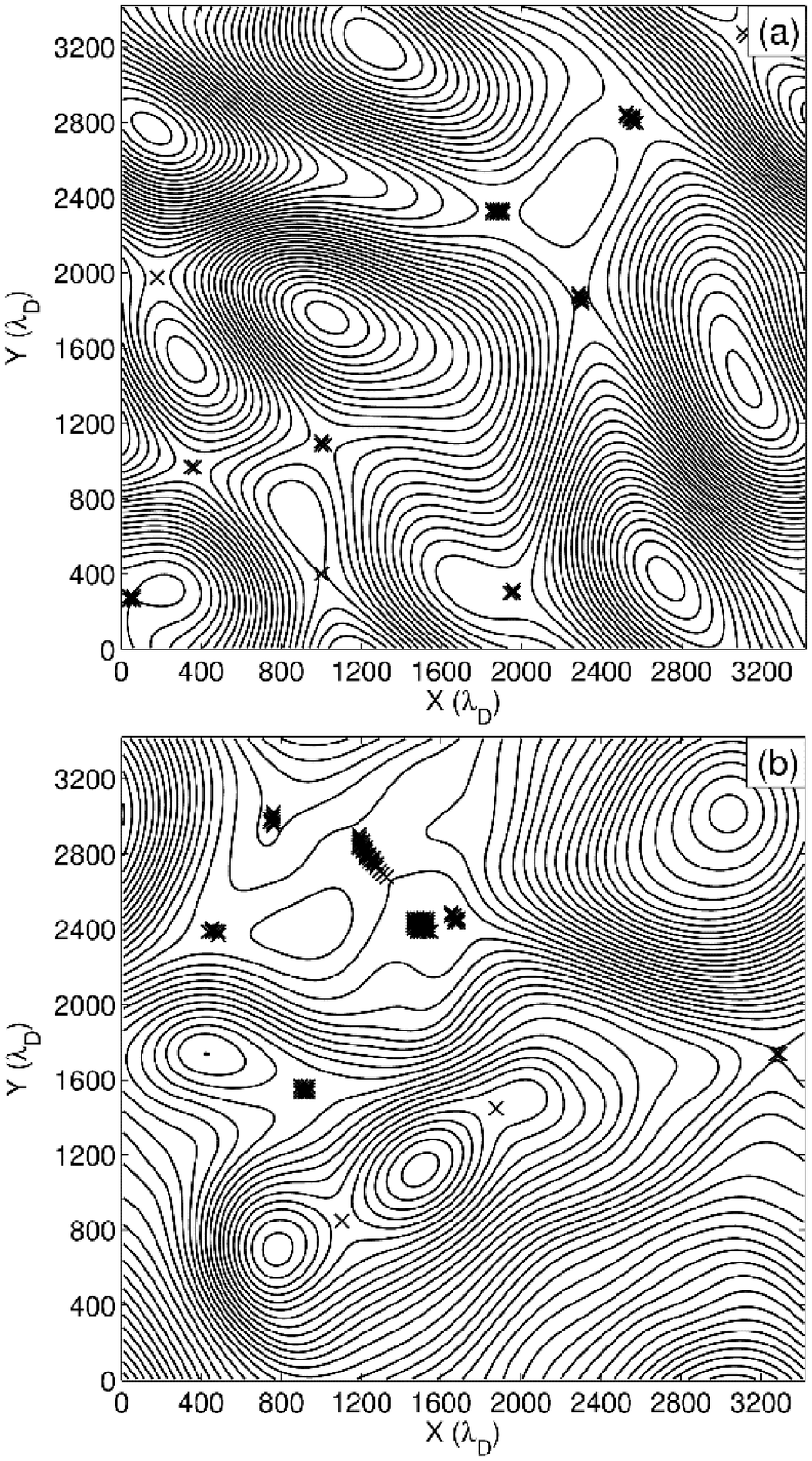} 
 \caption{ \label{fig:initialB2}
Initial and final magnetic field line configurations at (a) $t = 0$ and (b) $t =  200\Omega_e^{-1}$. Lengths in units of $\lambda_D$.}
\end{figure}

Figure~\ref{fig:initialB2} shows magnetic field line configurations for the simulation at times $t = 0$ and $t =  200\Omega_e^{-1}$. Potential X-point reconnection sites are shown by black crosses where the threshold for the gradient is satisfied. Sometimes a single X-point will be shown as a cluster of cells satisfying the magnetic geometrical criterion. The end state has a magnetic field which is topologically simpler with 8 X-point sites compared to 10 initially. When the simulation is run for longer times, beyond $t = 200\Omega_e^{-1}$, the number of X-points reduces further, with a consequent reduction in the number of magnetic islands. During the simulation the field line evolution is highly dynamic, with X-points moving around the simulation region and interacting with magnetic islands and other X-points. Animations show that magnetic islands gain or lose flux via reconnecting field line motion through X-points, i.e., field lines with island-like connectivity become linked to other X-points or encircle more than one island. Thus an island may shrink until it disappears or is absorbed by another island, and at this point the separating X-point also disappears. Occasionally new X-points are seen to form, but only temporarily as the field fluctuates. We also see that the sense of reconnection may reverse at an X-point, with the movement of field lines changing direction as the surrounding islands shrink or grow. Associated with a change in the sense of magnetic field reconnection there will also be a change in the sense of the reconnection electric field. In the final state there are some X-points which are in complex geometries and seem on the verge of disappearing if the simulation were run longer. By examining the time evolution of the local electron gyroradius we find that once an X-point region develops a scale less than about an electron gyrodiameter it is likely that the X-point will disappear, although sometimes disappearance occurs in more topologically complex regions with several X-points close to each other.

\begin{figure}[h]
\includegraphics[scale=0.53,angle=0]{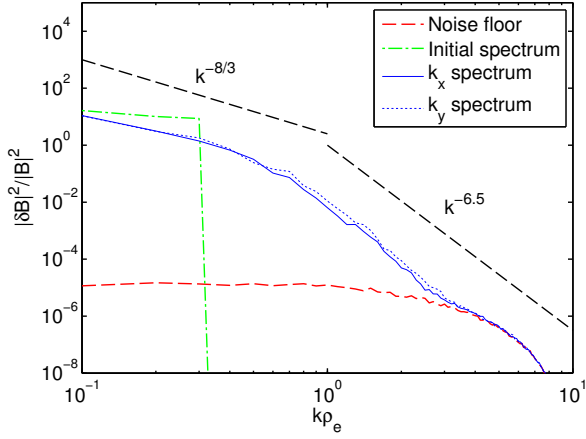}
\caption{\label{fig:powerspectra_kxky} Power spectra of $|\delta B|^2/|B_0|^2$ as functions of $k_x$ and $k_y$, at $t = 200\Omega_e^{-1}$ (blue solid and dotted lines), initial spectrum (green dash-dot line) and noise floor (red dashed line). A $k^{-8/3}$ gradient is shown to indicate the typical gradient seen in other works where $k$ extends to smaller values.}
\end{figure}

Figure~\ref{fig:powerspectra_kxky} shows power spectra of $|\delta B|^2/|B_0|^2$ as functions of $k_x$ and $k_y$ at $t =  0$ (green) and $t = 200\Omega_e^{-1}$ (blue). The $k_x$ and $k_y$ directions are both perpendicular to the average guide field. The noise floor, as determined from a simulation with no applied perturbation, is shown in red. Starting from the initial energy input at small $k$ values, the power at larger $k$ evolves rapidly until $t = 100\Omega_e^{-1}$, after which the spectra are relatively time steady over the period simulated. There are no major differences between the spectra in the two directions. Simulations with an in-plane guide field show a similar rapid formation of approximately power law spectra, but with a power anisotropy in the parallel and perpendicular wave vector directions \citep{camporeale:2011}. The simulation domain size is approximately one ion inertial length $\lambda_i$, and between this scale and $k\rho_e=0.3$ the power spectrum is approximately proportional to $k^{-8/3}$ in agreement with other simulations and observational data \citep{Smith:2006,Alexandrova:2008}. Beyond this driving scale, the spectra gradually steepens, roughly consistent with observations, until it reaches a power law of approximately $k^{-6.5}$, but there appears to be no obvious break point in wavenumber. Note that for $k\rho_e \geq 4$ the spectrum is not above the background noise level and is not meaningful.

In addition to the B spectrum shown in Fig.~\ref{fig:powerspectra_kxky} we have also examined the spectra of the electric field, number density and average ion and electron velocity. These spectra are qualitatively similar, showing a dual slope form of power law. These spectra also evolve until $t = 100\Omega_e^{-1}$, and show that an ensemble of stochastic fluctuations is present within the simulation, and has properties that are similar to turbulence. This indicates the formation of a turbulent cascade down to the noise level of the simulation and scales of order the electron gyroradius.

We have also run the same simulation with different sets of random initial perturbations in the magnetic field, and find qualitatively the same results as presented in the following sections, just with different topological configurations. Additionally, we have found that running the same simulation but with $m_i/m_e=400$ does not significantly change the form of the power spectra, the temperature, or temperature anisotropy signatures that we present below.

\subsection{Reconnection signatures}

Reconnection is usually understood to produce plasma heating, and so one might expect to find increased electron temperature around magnetic field X-points. However, over the course of the simulation, the changes in electron temperature seen in the vicinity of the reconnection sites are not significantly different from those at other locations. This indicates that energy dissipation occurring  through the reconnection process does not dominate over dissipation elsewhere. The lack of a unique, strong correlation between $T_e$ changes and the location of reconnection sites may be related to recent observations of magnetopause reconnection outflows that show a wide variability in bulk electron heating, explained by a dependence on the Alfv\'en speed of the inflow \citep{Phan:2013}. Consequently, in order to illustrate the effects of reconnection around magnetic field X-points, we concentrate on the electron drift velocity and temperature anisotropy.

\begin{figure}[hc]
\includegraphics[scale=0.38,angle=0]{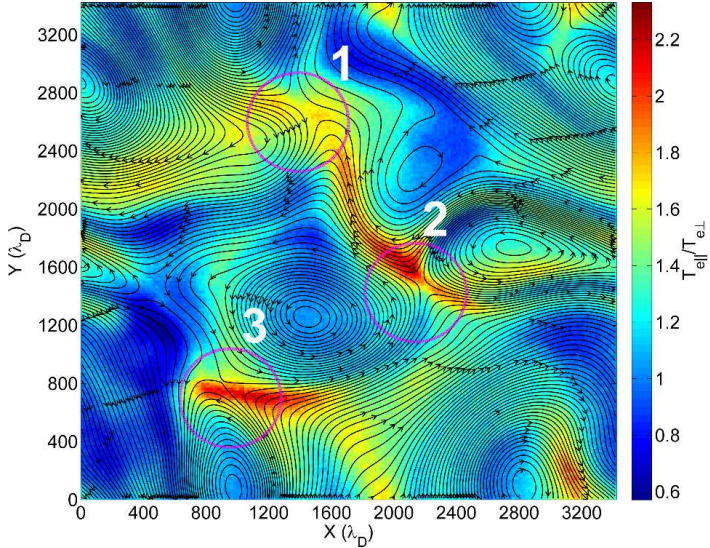}
\caption{\label{fig:3_drifts} Magnetic field lines (black) and electron temperature anisotropy $(T_{e\|}/T_{e\perp})$ at time $t = 97\Omega_e^{-1}$ for the full simulation domain. The three X-point regions discussed in the text are marked.}
\end{figure}

Figure~\ref{fig:3_drifts} shows the electron temperature anisotropy $T_{e\|}/T_{e\perp}$ over the full simulation domain at $t =  97\Omega_e^{-1}$, with magnetic field lines shown in black. Although $T_{e}$ does not significantly increase at the reconnection sites, three reconnection sites have been labeled where there is a strong signature of parallel temperature anisotropy. Our analysis will focus on these three sites in order to understand how this particular signature arises, and how it is associated with magnetic reconnection.

\begin{figure}[h]
\includegraphics[scale=0.4,angle=0]{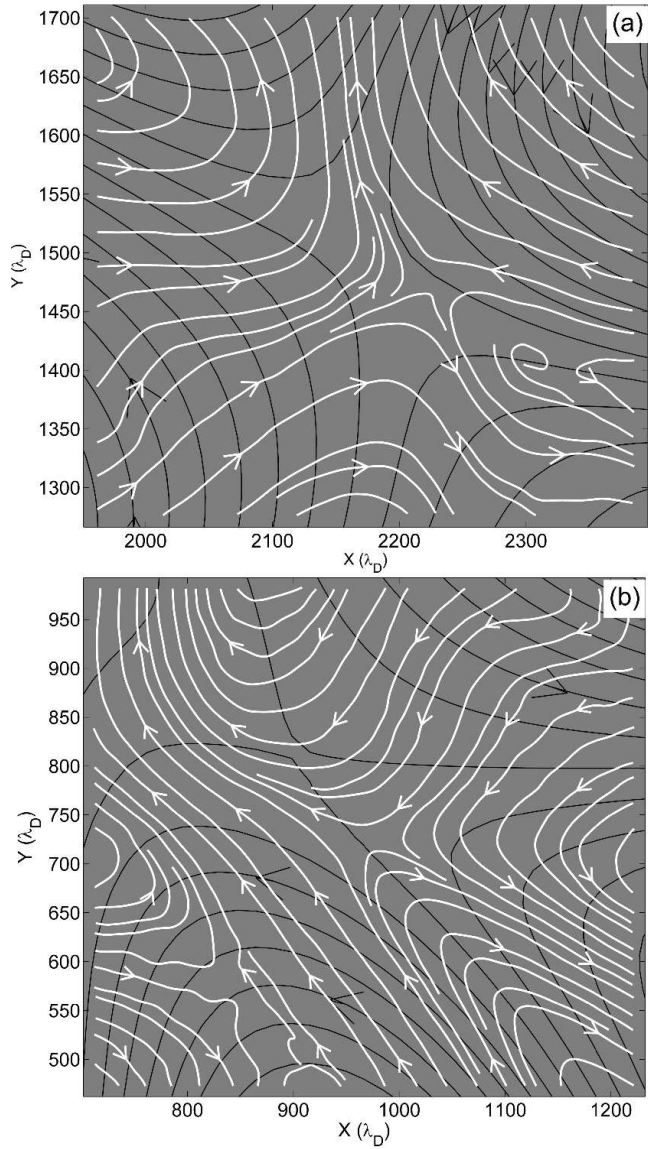}
\caption{\label{fig:electron_drifts} 
Enlarged detail showing magnetic field lines (black) and electron streamlines (white) for (a) reconnection site 2, and (b) reconnection site 3. The position of the reconnection sites are marked in Fig.~\ref{fig:3_drifts}.}
\end{figure}

Figure~\ref{fig:electron_drifts} shows an enlarged detail for reconnection sites 2 and 3. Electron average velocity streamlines are shown in white and magnetic field lines in black. The geometry of the magnetic field lines and electron flows in Fig.~\ref{fig:electron_drifts} resemble typical X-point reconnection with inflow and outflow regions. The regular flow configuration is perhaps surprising at this scale, given that the size of the reconnection site is $\sim 400 \lambda_D$ (25 cells), compared to the electron thermal gyroradius of $\sim 80 \lambda_D$. The velocity and the magnetic field patterns are not symmetric, and for site 2 the flow and field pattern centers are displaced from one another by $\sim 50 \lambda_D$. The corresponding displacement is larger for site 3, possibly due to the larger asymmetry imposed by the surrounding islands. Reconnection sites 2 and 3 are both at the junction of merging magnetic islands. Reconnection site 1 is located within a more complex magnetic topology, and the sense of field line motion changes as the local islands around it disappear. Despite this, this site shows a large electron parallel temperature anisotropy and an electron flow signature similar to the other two sites.

\begin{figure}
\includegraphics[scale=0.4,angle=0]{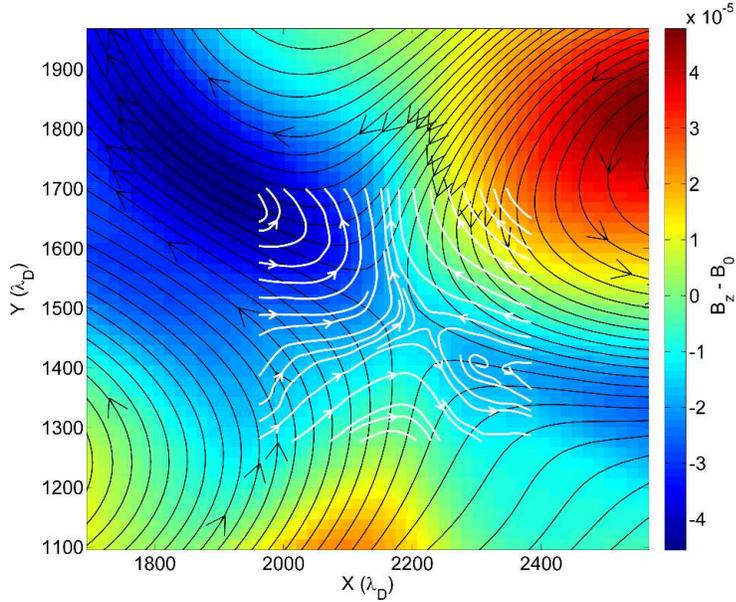}
\caption{\label{fig:quadrupole} Out of plane magnetic field component ($B_z - B_0$) showing quadrupolar signature around reconnection site 2. Magnetic field lines shown in black, and electron streamlines in white.}
\end{figure}

Figure~\ref{fig:quadrupole} again shows reconnection site 2, with field lines plotted in black, electron streamlines plotted in white, and $B_z - B_0$ plotted as a color map. Subtracting the initial guide field from $B_z$ reveals the shape of the out-of-plane quadrupolar signature, as seen in two-fluid Hall MHD simulations \citep{Sonnerup:1979,Terasawa:1983,Karimabadi:2004}, hybrid simulations \citep{Karimabadi:1999}, and full particle simulations simulations \citep{Lapenta:2011}. This quadrupolar signature arises from the circular motion of electron currents in the region, as they decouple from the ion flow, which enhances the out-of-plane magnetic field. Anticlockwise electron motion creates a negative enhancement in the magnetic field, as can be seen in the top-left and bottom-right of Fig.~\ref{fig:quadrupole}, whereas clockwise electron motion creates a positive enhancement, as at top-right and bottom-left. The asymmetry of the X-point configuration is also seen in the asymmetric quadrupolar Hall signature. It is known that the presence of a guide field can result in an asymmetric reconnection field pattern due to the nonlinear interaction between guide field and Hall field components \citep{Karimabadi:1999,Eastwood:2010}. However, asymmetry can be caused by other factors, such as density gradients and asymmetric in-flows driving the reconnection, both of which are present in this simulation. It is due to these signatures that the reconnection shown can be described as Hall reconnection and is mainly due to the interaction of decoupled electron and ion flows. The reconnection arises spontaneously, driven by the plasma dynamics introduced by the initial magnetic perturbation.


Figure~\ref{fig:aniso_site2} shows the distinctive shape of the region of increased electron temperature anisotropy generated around reconnection site 2. Two main areas of strong temperature anisotropy are located to the top-left and the bottom-right of the center of reconnection in the outflow regions. An animation of the time development of the temperature anisotropy and magnetic field lines shows that the anisotropy increases with the reconnection rate and appears to grow out from the center of reconnection. (See animation of Fig.~\ref{fig:aniso_site2} online.)

In a 2-D geometry the rate of reconnected flux, i.e., the reconnection rate, is equal to the out of plane electric field $E_z$, since
\begin{equation} \label{eq:Recon1}
\textbf{E} = - \nabla V - \frac{\partial \textbf{A}}{\partial t},
\end{equation}
where \textbf{A} is the magnetic vector potential and $V$ is the electrostatic potential. Field lines in 2-D are contours of constant $A_z$, so the rate of in-plane reconnection depends only on $A_z$. With $z$ as the ignorable coordinate, it follows that:
\begin{equation} \label{eq:Recon2}
E_z = - \frac{\partial A_z}{\partial t},
\end{equation}
so that $E_z$ at the center of a reconnection site  corresponds to the reconnection rate.

The centers of the identified reconnection events were tracked during their motion in the course of the simulation and $E_z$ recorded, as also were parameters such as electron temperature and anisotropy. Parameter values were averaged over a box size of 15 cells square, centered on the reconnection site. Measured reconnection rates were consistent with the animations of magnetic field line motion, further evidence that reconnection was occurring. Figure~\ref{fig:ReconRate} shows time series of the resulting data for reconnection site 2. Figure~\ref{fig:ReconRate}(a) shows reconnection rate, Fig.~\ref{fig:ReconRate}(b) shows average electron temperature and Fig.~\ref{fig:ReconRate}(c) shows average electron anisotropy in terms of the parameter $(1-T_{e\|}/T_{e\perp})$. In this figure an anisotropy parameter value less than zero corresponds to a parallel temperature anisotropy.


As previously noted, the electron temperature is not distinctly greater around reconnection sites when compared with other temperature variations in the simulation. Comparing the time profiles of Fig. \ref{fig:ReconRate}, $T_e$ varies by $\pm2$ \% of its initial value, before $t = 100\Omega_e^{-1}$. There is a small correlation with reconnection rate, but after this time the electrons around the reconnection site experience overall cooling and the correlation becomes weak. However, a correlation is evident between anisotropy and reconnection rate, throughout the entire simulation, indicating that the reconnection process is responsible. Not all reconnection sites show such a marked increase in anisotropy, which suggests a more complex physical process is operating rather than a simple, local one which would produce an absolute correlation with reconnection rate.

We have examined the ion distribution functions in the vicinity of the reconnection site, and there are no significant changes in ion temperature or temperature anisotropy. The ion distribution functions remain isotropic. This is not unexpected considering the minor role of the ion dynamics over the timescale of the simulation.

\subsection{Electron velocity distribution functions}

\begin{figure}
\includegraphics[scale=0.36,angle=0]{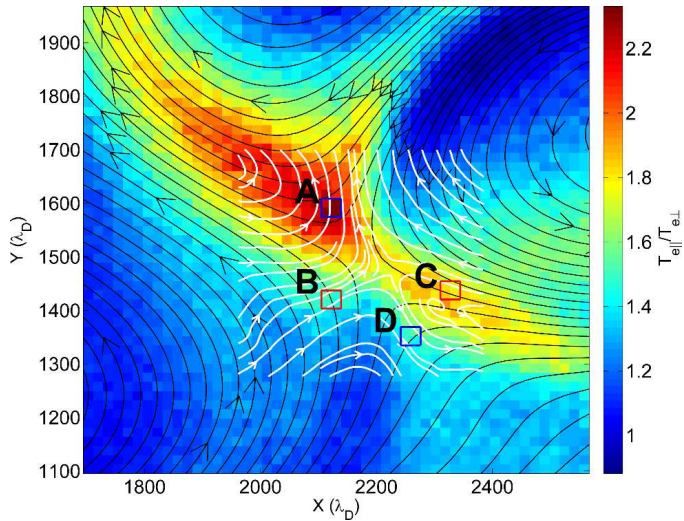}
\caption{\label{fig:aniso_site2} Electron temperature anisotropy ratio $T_{e\|}/T_{e\perp}$ in the region of reconnection site 2. Magnetic field lines are shown in black and electron streamlines are shown in white. The four marked regions are discussed in the text. This figure is available as an animation in the electronic edition of the {\it Astrophysical Journal}. The animation shows the time development of $T_{e\|}/T_{e\perp}$ and motion of field lines through the reconnection site.}
\end{figure}
\begin{figure}
\includegraphics[scale=0.75,angle=0]{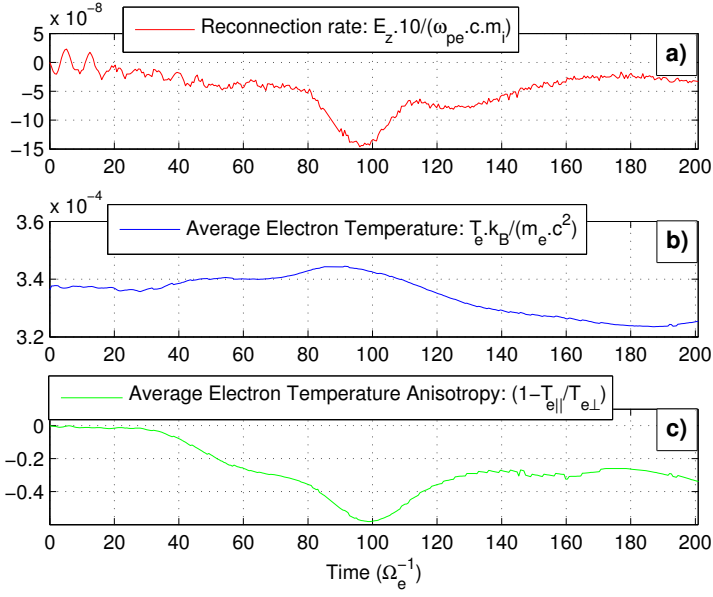}
\caption{\label{fig:ReconRate} Averaged values around reconnection site 2 for (a) reconnection rate, (b) electron temperature, and (c) temperature anisotropy parameter $(1-T_{e\|}/T_{e\perp})$ as a function of time.}
\end{figure}
\begin{figure*}[t]
\includegraphics[scale=0.59,angle=0]{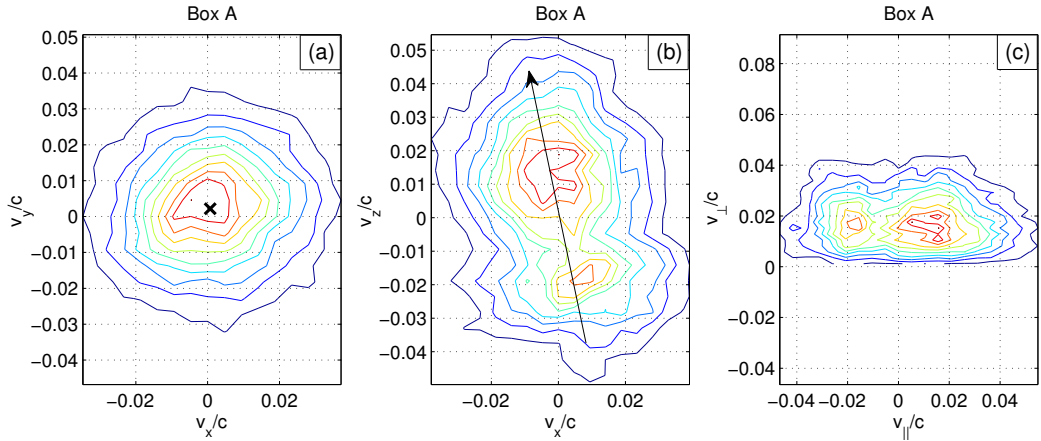}
\caption{\label{fig:VDF_a} Electron VDF for reconnection site 2, box A (cf. Fig.~\ref{fig:aniso_site2}) for (a) $v_x-v_y$, (b) $v_x - v_z$, and (c) $v_\| - v_\perp$ planes.}
\end{figure*}

We now discuss the velocity distributions seen near reconnection site 2 shown in Fig.~\ref{fig:aniso_site2}. Four sub-regions are chosen on different sides of the magnetic separatrices corresponding to electron in-flow (B and C) and electron out-flow (A and D). Figure~\ref{fig:VDF_a} shows the electron distribution in the region of inflow box A, where the anisotropy is highest. Figure~\ref{fig:VDF_a}(a) shows the distribution in the $v_y - v_x$ plane, with a black cross at the electron bulk velocity; Fig.~\ref{fig:VDF_a}(b) shows the distribution in the $v_z - v_x$ plane, with the black arrow indicating the direction of the magnetic field; and Fig.~\ref{fig:VDF_a}(c) shows the distribution in the $v_\perp$ - $v_\|$ plane. All distributions are normalized to unity with red contours indicating higher particle density than blue; velocities are normalized to the speed of light $c$.

From Fig.~\ref{fig:VDF_a}(c) it can be seen that thermal width of the distribution in the parallel direction is approximately 1.5 times that in the perpendicular direction, in agreement with the anisotropy ratio of approximately $2.3$ (cf. Fig.~\ref{fig:aniso_site2}). Rather than a bi-Maxwellian shape, the distribution shows a double peaked, beam-like structure, with one peak on the negative $v_\|$ side and another on the positive $v_\|$ side of the distribution. The drift velocity in the positive $y$ direction is consistent with the flow pattern of Fig.~\ref{fig:aniso_site2}. The symmetry of the $v_z$~-~$v_x$ distribution around the magnetic field direction seen in Fig.~\ref{fig:VDF_a}(b) indicates that the electron distributions are approximately gyrotropic.

\begin{figure*}[t]
\includegraphics[scale=0.59,angle=0]{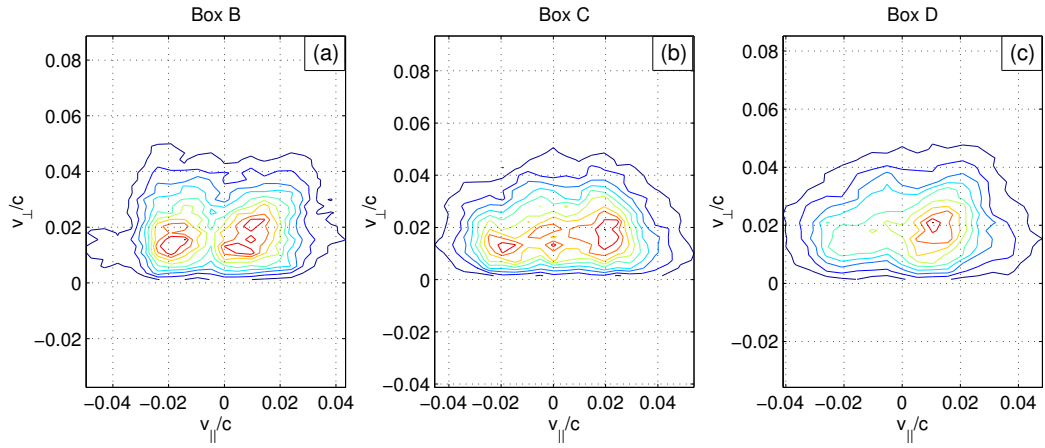}
\caption{\label{fig:VDF_bcd} Electron VDF in $v_\| - v_\perp$ plane for reconnection site 2, for (a) box B, (b) box C and (c) box D, as marked in Fig.~\ref{fig:aniso_site2}.}
\end{figure*}

Figure~\ref{fig:VDF_bcd} shows plots of the $v_\perp$ - $v_\|$ distribution functions for the other three boxes B, C, and D marked in Fig.~\ref{fig:aniso_site2}. These all show multi-peaked structures. For example, Fig.~\ref{fig:VDF_bcd}(c) box D has the appearance of a core plus beam distribution, as the peak on the positive $v_\|$ side of the distribution is much larger. The distribution for box C (Fig.~\ref{fig:VDF_bcd}(b)) even shows a triple peaked distribution. Thus the regions of largest temperature anisotropy in Fig.~\ref{fig:3_drifts}, which occur around reconnection sites, seem to correspond to the presence of distribution functions with a mix of multiple peaks. This, in itself, suggests that the reconnection process is forming one or both of these peaks, possibly by accelerating a subset of particles to form a second peak. 

\subsection{Particle Tracking}

In order to determine the formation mechanism of these multi-peaked distributions, we track particle trajectories of electrons selected from the different peaks of the distributions, to determine whence these separate populations of electrons originate. Although the complex physics cannot be understood merely in terms of single particle motions, this exercise will allow us to show whether one or both of the distribution function peaks have been produced by electrons being accelerated or decelerated as they approach and interact with the reconnection site.

\begin{figure}
\includegraphics[scale=0.7,angle=0]{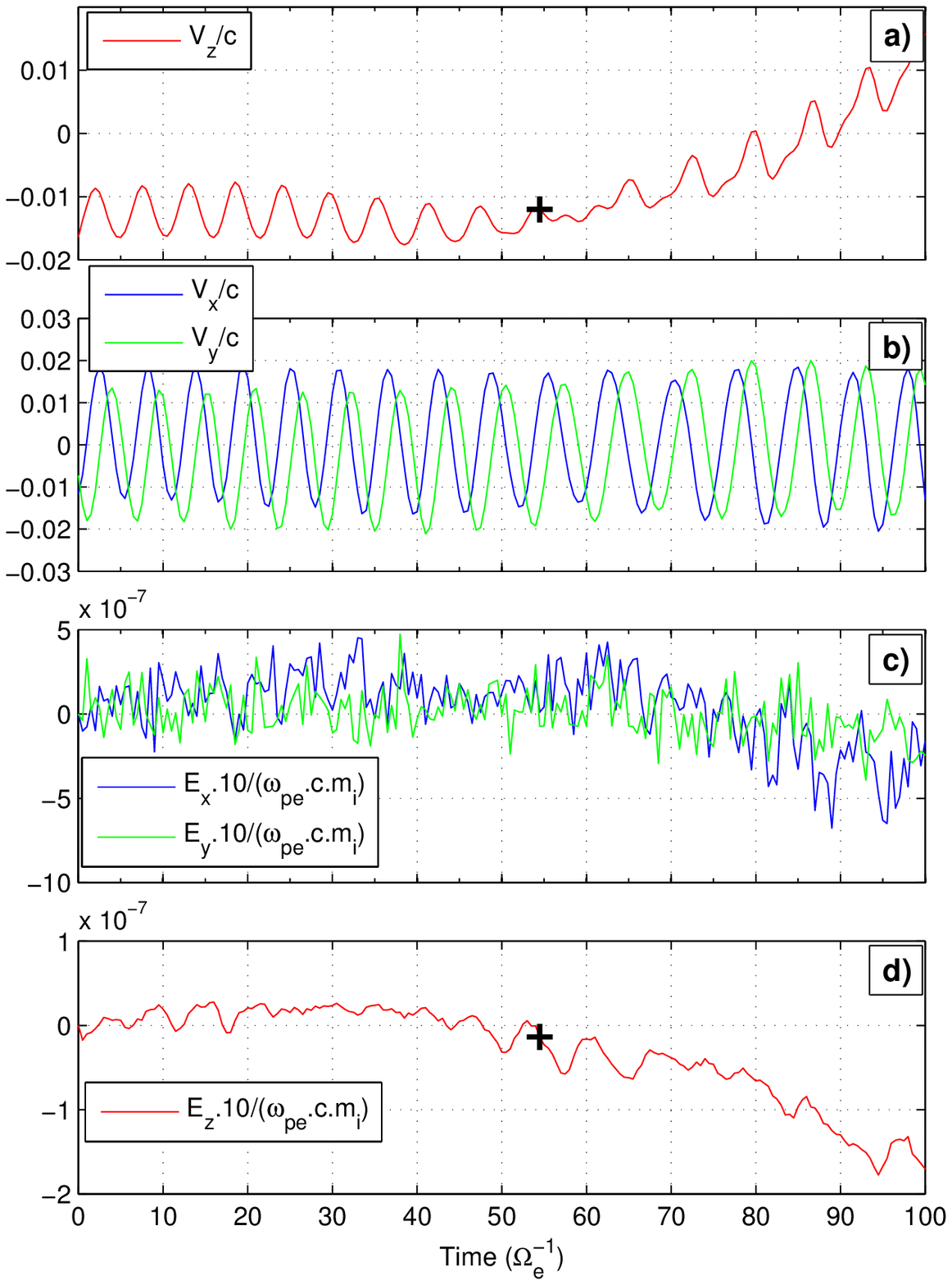}
\caption{\label{fig:Accel1} Time series of particle velocity components and electric field components (as experienced by the particle) for electron E1 which is chosen from the positive $v_\|$ peak in the distribution function for box A (Figs.~\ref{fig:aniso_site2}, \ref{fig:VDF_a}).}
\end{figure}

\begin{figure}

\includegraphics[scale=0.68,angle=0]{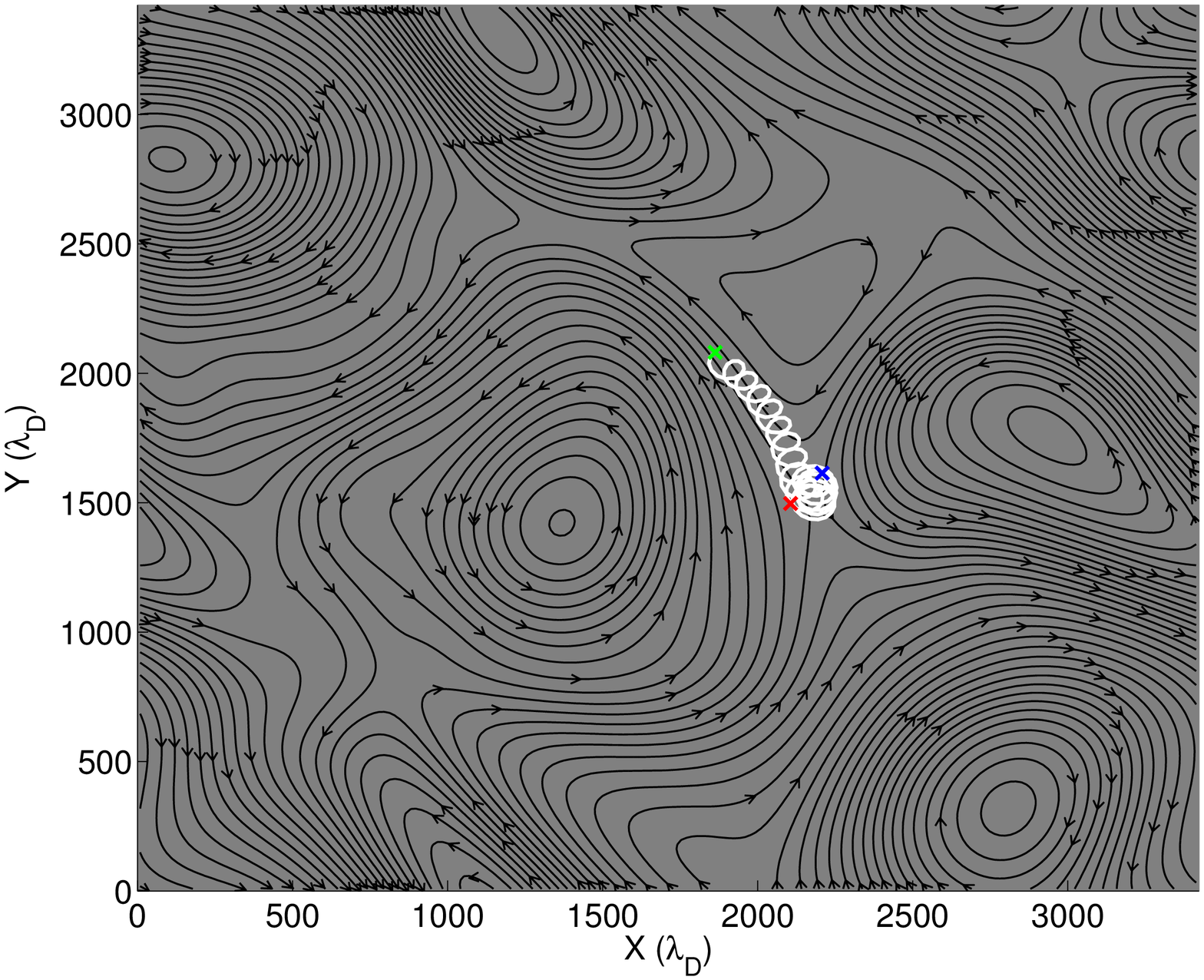} 
\caption{\label{fig:Trajectory1} Trajectory (white) of electron E1 as it approaches and interacts with reconnection site 2. The start and end locations are shown by green and blue crosses respectively. Magnetic field lines (black) are plotted at the time indicated by the black cross in Fig.~\ref{fig:Accel1}. The position of electron E1 at the time of the plotted field lines is shown with a red cross.}
\end{figure}

\begin{figure}

\includegraphics[scale=0.7,angle=0]{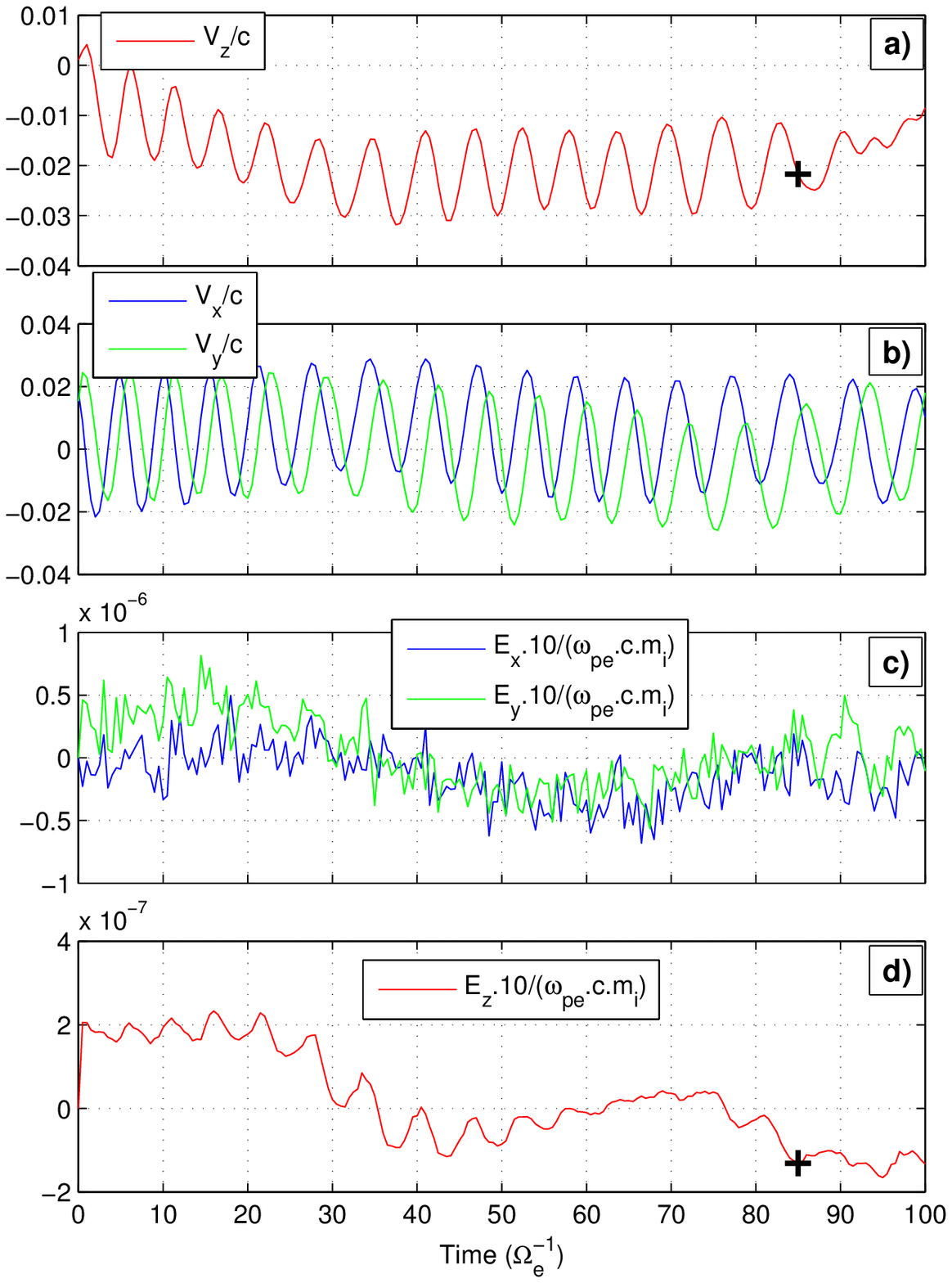}
\caption{\label{fig:Accel2} Time series of particle velocity components and electric field components (as experienced by the particle) for electron E2 which is chosen from the negative $v_\|$ peak in the distribution function for box A (Figs.~\ref{fig:aniso_site2}, \ref{fig:VDF_a}).}
\end{figure}

\begin{figure}
\includegraphics[scale=0.68,angle=0]{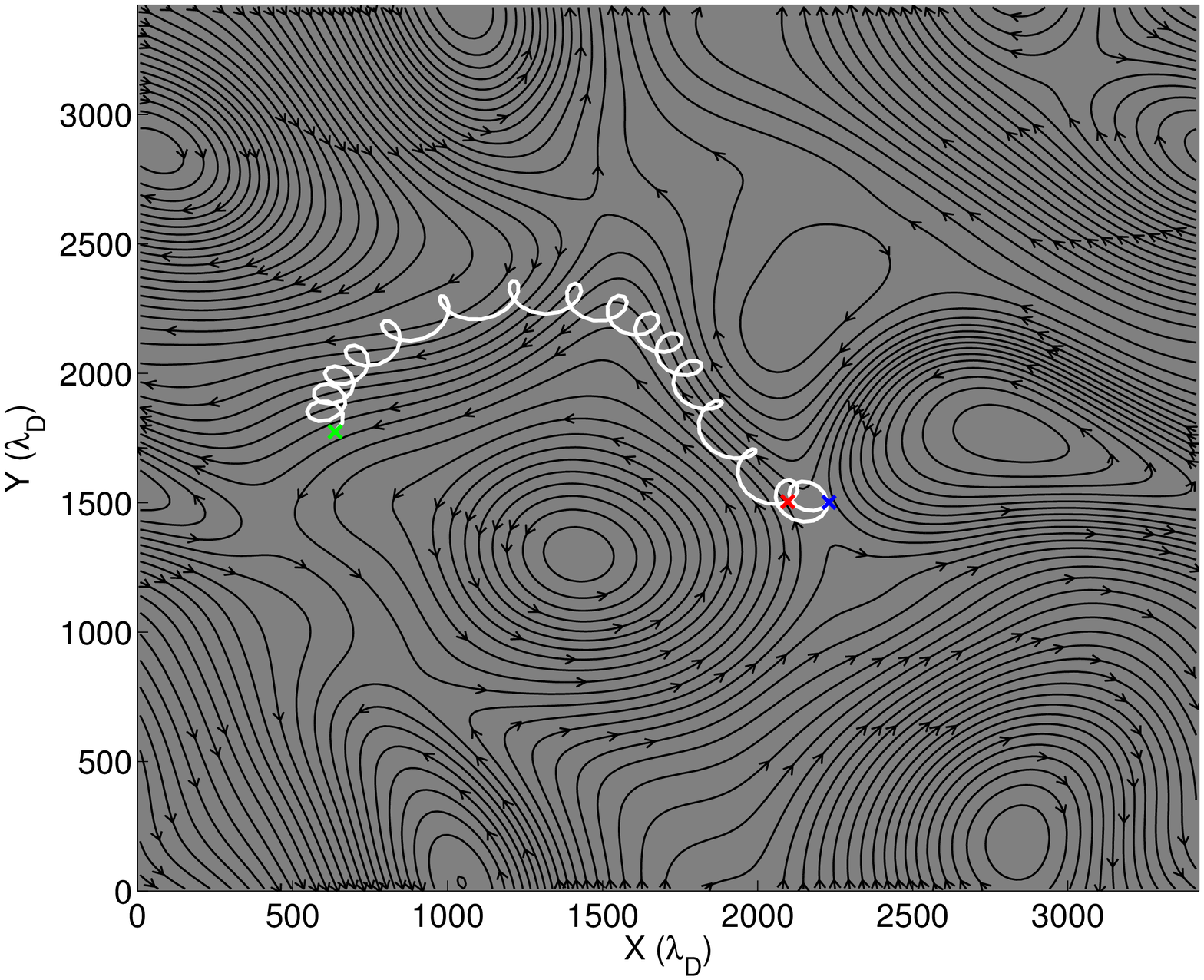} 
\caption{\label{fig:Trajectory2} Trajectory (white) of electron E2 as it approaches and interacts with reconnection site 2. The start and end locations are shown by green and blue crosses respectively. Magnetic field lines (black) are plotted at the time indicated by the black cross in Fig.~\ref{fig:Accel2}. The position of electron E2 at the time of the plotted field lines is shown with a red cross.}
\end{figure}

We show data for two electrons, labeled E1 and E2, which were tracked throughout the simulation. Both electrons were located in box A (Fig.~\ref{fig:aniso_site2}), and were chosen from a large set of recorded particles that interacted with reconnection site 2. Electron E1 (Figs.~\ref{fig:Accel1} and \ref{fig:Trajectory1}) was chosen from the particles in the peak on the positive $v_\|$ side of the distribution shown in Fig.~\ref{fig:VDF_a}(c). Electron E2 (Figs.~\ref{fig:Accel2} and \ref{fig:Trajectory2}) was chosen from the particles in the peak on the negative $v_\|$ side of the distribution. In these figures time has been normalized to $\Omega_e$ calculated using magnetic field $\mathbf{B_0}$. Fig.~\ref{fig:Accel1}(a) shows $v_z$ versus time for electron E1. Velocity components $v_x$ and $v_y$ are shown in Fig.~\ref{fig:Accel1}(b) in blue and green, respectively. The electric field as experienced by electron E1 is shown in the Fig.~\ref{fig:Accel1}(c) and (d), with $E_x$ and $E_y$ plotted in blue and green, and $E_z$ in red.

Figure~\ref{fig:Trajectory1} shows the trajectory taken by electron E1 before and during its encounter with the reconnection site. Magnetic field line contours for the whole simulation box are shown in black at $t = 55 \Omega_e^{-1}$. Traced in white is the electron position for the interval $t = 0$ to $t = 100 \Omega_e^{-1}$. The green and blue crosses on the trajectory mark the start and end locations, respectively. In Fig.~\ref{fig:Accel1} the black crosses marked on the $v_z$ and $E_z$ time-series, corresponds to the time at which the magnetic field lines are shown, with a corresponding red cross marked on the electron trajectory at the same time (Fig.~\ref{fig:Trajectory1}). It is important to remember that the magnetic field evolves dynamically over the time interval of the electron trajectories. Thus, the magnetic field line configuration shown in these figures is only illustrative of the magnetic environment at a specific time late in the trajectories. Animations have been used extensively to analyze the electron trajectories relative to the dynamic field line geometry.

Figure~\ref{fig:Accel1}(a) shows that as electron E1 encounters the reconnection region after $t = 55 \Omega_e^{-1}$, the parallel $z$ component of its velocity increases. This is one example of many particles that were tracked, and all show similar behavior. Electrons experience an acceleration, due to $E_z$, along the guide field direction as they approach the reconnection site. The large increase in negative $E_z$ in Fig.~\ref{fig:Accel1}(d) results in a force on electron E1 in the positive $z$ direction; $E_z$ at reconnection site 2 is mainly negative throughout the simulation (Fig.~\ref{fig:ReconRate}(a)). Figure~\ref{fig:Accel1}(a) also indicates that electron E1 passes very close to the center of reconnection, since the oscillation in $v_z$ decreases in amplitude, indicating that the in-plane components of the magnetic field have become almost zero. In summary, reconnection site 2 is responsible for the positive parallel peaks in Figs.~\ref{fig:VDF_a} and \ref{fig:VDF_bcd}.


Figures~\ref{fig:Accel2} and \ref{fig:Trajectory2}, in the same format as Figs.~\ref{fig:Accel1} and \ref{fig:Trajectory1} respectively, show trajectory information for electron number E2, which is from the peak on the negative $v_\|$ side of the distribution of Fig.~\ref{fig:VDF_a}. This particle has experienced acceleration in the negative $z$ direction before it encounters the reconnection site at approximately $t = 85 \Omega_e^{-1}$. However, it, like electron E1, experiences a positive acceleration after it enters the region around reconnection site 2, consistent with the negative $E_z$. However, despite this acceleration the particle $v_z$ remains negative. We have examined 181 particle trajectories taken from the negative $v_\|$ peak of the distribution and they all show a similar history; there are a total of 9901 simulation particles in the distribution with $v_\|<0$. From Figs.~\ref{fig:Trajectory1} and~\ref{fig:Trajectory2} electrons E1 and E2 have very different trajectory histories, but are eventually colocated but with very different parallel velocities.

Since reconnection site 2 has mainly negative $E_z$ for most of the simulation, it accelerates electrons in the positive $z$ direction. We have used this fact to confirm this mechanism of temperature anisotropy generation, by examining the positions at previous time-steps of groups of particles from the negative peaks of Figs.~\ref{fig:VDF_a} and \ref{fig:VDF_bcd}. The trajectories of these particles trace a region whose shape is towards the center of reconnection only from the top left part of the separatrix and the bottom right part of the separatrix, consistent with the shape of the region of enhanced parallel anisotropies in these locations (Fig.~\ref{fig:aniso_site2}).

In order to determine why a triple peaked distribution is formed, as shown in box C (Fig.~\ref{fig:VDF_bcd}), particles from the central peak were also tracked. Although not shown here, these particles again show positive increases in $v_z$ near the reconnection site. So the central peak is formed of particles that start with a negative $v_z$ but as they enter the reconnection site they are only accelerated enough to finish in the center of the distribution. So double or triple peaked distributions can be formed by electrons with different trajectory histories passing through multiple acceleration regions, but arriving at the same location within a reconnection site.

\section{Conclusion}

We have presented the results of 2-D simulations using realistic proton to electron mass ratio of the turbulent decay of large scale fluctuations with an out-of-plane guide field. As in previous similar work with the guide field in the simulation plane \citep{camporeale:2011} a fluctuation power spectrum with approximately power law form quickly evolves, until $t = 100\Omega_e^{-1}$, after which the spectra are relatively time steady over the period simulated. The spectra extends to small scales of order the electron gyroradius. Animations of the magnetic field evolution show that X-points (i.e., potential reconnection sites) evolve dynamically, responding to the motion of surrounding magnetic islands in the turbulence. As reconnection occurs the topology of field lines can change as they move through the X-points, from closed within a single magnetic island to circulating around several islands. The sense and rate of field line motion can change at any one particular X-point as the islands surrounding it grow or shrink. During the course of the simulation a number of the initial X-points disappear, and this is most likely to happen after the scale of the X-point becomes less than the local electron gyrodiameter. When the simulation is run for longer times the number of X-points reduces further, with a consequent reduction in the number of magnetic islands. Thus the simulation sees a relaxation of the initial magnetic topology, as well as a redistribution of power from large to short scales.

The regions around X-points have signatures which indicate that magnetic reconnection is occurring, with the motion of field lines and the pattern of electron bulk drifts consistent with reconnection inflows and outflows. Where there is a clear pattern of reconnection associated electron drifts we also observe a quadrupolar signature in $B_z$, similar to that found in Hall reconnection. This is consistent with the scale of the X-point region being smaller than the ion inertial and gyro-scales, so that the electron and ion motion are effectively decoupled. Generally there are asymmetries in the quadrupolar signature and flow pattern due to the guide field, density gradients and inflows. Because of the size of the simulation (the largest scale is of order the ion inertial length) and the initial number and shape of the islands, we do not see the formation of narrow (small aspect ratio) current layers with embedded X-points. A larger simulation, with initial fluctuation injection at larger scales, or with initial power anisotropy may produce a different geometry of initial X-points in narrow current sheets as seen in MHD simulations \citep{servidio:2009} and some PIC simulations \citep{Karimabadi:2013}

Animations of the evolution of the electron temperature anisotropy ratio $T_{e\|}/T_{e\perp}$ indicate enhanced parallel anisotropy at some X-points, and the dynamic appearance of regions of enhanced parallel anisotropy in reconnection outflow regions during periods of strong reconnection. There is not a unique one-to-one correspondence between X-points and regions of enhanced anisotropy, but this behavior is frequently observed. We have shown that the enhanced anisotropy is due to multi-peaked velocity distribution functions. This is the first time (to our knowledge) that such velocity space structures have been reported at this scale and in turbulence. Further investigation reveals that such distributions are not unique to reconnection outflow regions, but can be found elsewhere in the simulation.

\begin{figure}
\includegraphics[scale=0.6,angle=0]{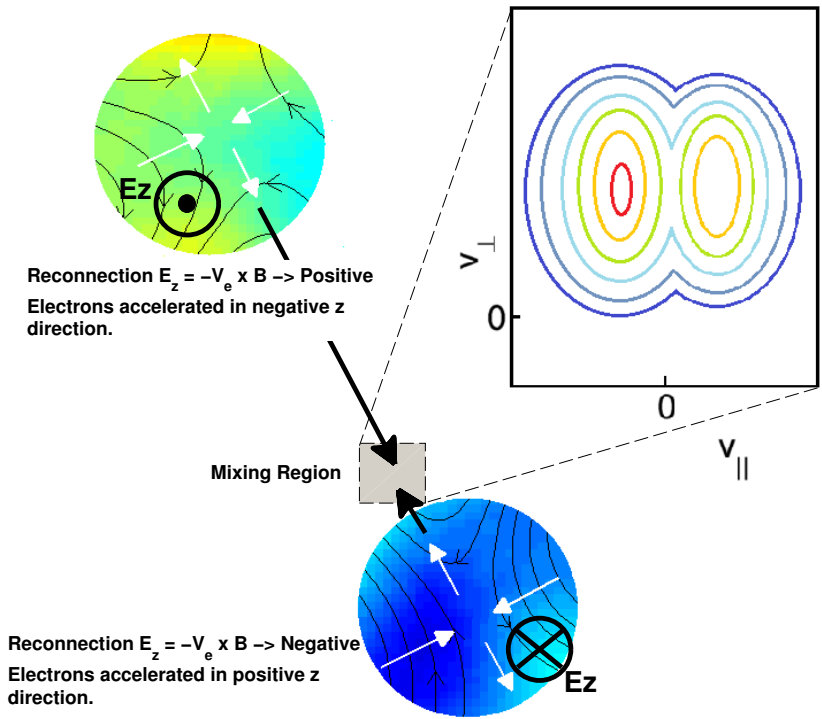} 
\caption{\label{fig:schematic} Schematic of mechanism for electron temperature anisotropy production due to reconnection in turbulence. Electrons accelerated at a region with a positive reconnection electric field $E_z$, gaining $v_\|<0$, can propagate along reconnected field lines towards another reconnection site with $E_z$ negative. Other electrons accelerated more locally gain $v_\|>0$, and the two populations form a double peaked distribution in a mixing region in the reconnection outflow.}
\end{figure}

In order to determine how these velocity space features are formed, and whether reconnection sites are responsible, electrons from the peaks of the distribution were tracked. It was found that electrons are accelerated  by the reconnection electric field $E_z$, in the direction of the guide field, when they are close to a reconnection site. Acceleration can occur in both positive and negative $z$ directions depending on the sense of reconnection at a particular X-point. Particle tracking allows us to give the following explanation of the mechanism (see Fig.~\ref{fig:schematic}): The main peak of the distribution is generated by the local reconnection site, with the direction being set by the sense of reconnection, i.e. the sign of $E_z$. The outflow of electrons with the shifted $v_z$ distribution will then potentially mix with the surrounding population of electrons. Large anisotropies therefore form around a reconnection site whose outflow area already has a population of electrons, accelerated near another reconnection site, shifted in $v_z$ in the opposite sense. This produces the double peaked distributions which are seen. This mechanism explains why not all reconnection sites in the simulation show this large temperature anisotropy signature, it depends on both the current direction of reconnection for the site, and the presence of a population of electrons oppositely shifted in velocity in its outflow region. This in turn depends on the magnetic topology of field lines allowing electron trajectories to connect different reconnection sites. In this model the reconnection sites act as both acceleration regions and mixing zones. It is also possible in this scenario to explain the presence of triple-peaked distribution functions, which are sometimes seen.

We expect the multi-peaked distributions may be unstable preferentially for parallel/oblique propagating waves, but given the guide field direction and 2-D nature of our simulation it is unlikely that the unstable waves are supported. In a full 3-D simulation we suggest that these multi-peaked distributions would produce additional waves via beam or anisotropy instabilities. It is not clear what the full effect of this would be in terms of electron scattering or magnetic field line topology, given that in a 3-D simulation the reconnection sites themselves would have their own three dimensional dynamics. In future work we will consider what type of instabilities and waves might be associated with these distributions, and their consequences.

The simulation results indicate that turbulence may play an active role in increasing electron parallel temperature anisotropy. This has implications for the study of the evolution of solar wind parameters which has highlighted the importance of kinetic linear instabilities in limiting temperature anisotropy in response to Coulomb collisions, and the expansion of the solar wind \citep{camporeale:2008, stverak:2008, Matteini:2012}. Our results indicate that reconnection can be another driver of electron temperature anisotropy. The simulation has possible limitations due to the size of the simulation box and the large amplitude of the initial fluctuations. These have been adopted due to constraints of realistic mass ratio, and the requirement to resolve a turbulent cascade above the noise floor of the simulation. Thus our results are more appropriate to, for example, the large amplitude turbulence behind the quasi-parallel terrestrial bow shock \citep{retino:2007} or in current sheets in the solar wind where some evidence of enhanced dissipation exists \citep{osman:2011}. 

Finally, the power spectrum of fluctuations that we observe develops rapidly after the start of the simulation, and has a power law form which is relatively time-steady. It does not seem directly influenced by reconnection, the dynamic behavior of X-points or the evolution of the electron temperature anisotropy.
A full analysis of the fluctuations contributing to the power spectrum and found during the relaxation process will be addressed in future work. Since the electron behavior is crucially dependent on the topological evolution of the magnetic field via reconnection, it seems possible that dissipation at the smallest scales in a collisionless plasma might be strongly influenced by how topological complexity is carried to small scales.



\acknowledgments

CTH is supported by an STFC (UK) studentship; DB is partially supported by STFC (UK) grant ST/J001546/1.




\bibliographystyle{apj}




\end{document}